# A new probabilistic shift away from seismic hazard reality in Italy?

A. Nekrasova[1,2], A. Peresan[2,3,5], V.G. Kossobokov[1,2,4,5], G.F. Panza[2,3,5,6]

[1] Institute of Earthquake Prediction Theory and Mathematical Geophysics, Russian Academy of Sciences, Moscow, Russian Federation
[2] The Abdus Salam International Centre for Theoretical Physics, SAND Group, Trieste – Italy
[3] Department of Mathematics and Geosciences, University of Trieste, Trieste, Italy. E-mail: aperesan@units.it
[4] Institut de Physique du Globe de Paris, Paris, France
[5] International Seismic Safety Organization (ISSO) - www.issoquake.org
[6] Institute of Geophysics, China Earthquake Administration, Beijing, People's Republic of China

**Abstract**

Objective testing is a key issue in the process of revision and improvement of seismic hazard assessments. Therefore we continue the rigorous comparative analysis of past and newly available hazard maps for the territory of Italy against the seismic activity observed in reality. The final Global Seismic Hazard Assessment Program (GSHAP) results and the most recent version of Seismic Hazard Harmonization in Europe (SHARE) project maps, along with the reference hazard maps for the Italian seismic code, all obtained by probabilistic seismic hazard assessment (PSHA), are cross-compared to the three ground shaking maps based on the duly physically and mathematically rooted neo-deterministic approach (NDSHA). These eight hazard maps for Italy are tested against the available data on ground shaking. The results of comparison between predicted macroseismic intensities and those reported for past earthquakes (in the time interval 1000 – 2014) show that models provide rather conservative estimates, which tend to over-estimate seismic hazard at the ground shaking levels below the MCS intensity IX. Only exception is represented by the neo-deterministic maps associated with a fixed return period of 475 or 2475 years, which provide a better fit to observations, at the cost of model consistent 10% or 2% cases of exceedance respectively. In terms of the Kolmogorov-Smirnov goodness of fit criterion, although all of the eight hazard maps differ significantly from the distribution of the observed ground shaking reported in the available Italian databases, the NDSHA approach appears to outscore significantly the PSHA one.

**Keywords** Earthquake catalogs – Peak ground acceleration – Macroseismic intensity – Probabilistic seismic hazard assessment – Neo-deterministic seismic hazard assessment – Seismic hazard maps

# 1. Introduction

A reliable and comprehensive characterization of expected seismic ground shaking, in a anticipatory perspective, is essential in order to develop effective risk mitigation strategies, including the adequate engineering design of earthquake-resistant structures.

A common belief is that a probabilistic assessment of the seismic hazard (PSHA), accounting for the probability of occurrence of a given ground shaking within a specified time interval, is needed for any rational decision making and for optimal allocation of resources (Marzocchi, 2013). However, since data are often insufficient to constrain the probability models and to test them, ground shaking probabilities turn out highly uncertain and unreliable, particularly for the large, sporadic and most destructive earthquakes. Comparison of observed numbers of fatalities with those calculated based on expected ground shaking from GSHAP maps (Wyss et al., 2012; Kossobokov and Nekrasova, 2012), show that seismic hazard maps based on the standard probabilistic method do not allow to reliably estimate the risk to which the population is exposed due to large earthquakes in many regions worldwide.

Although testing should be a necessary step in any scientific process of seismic hazard assessment, it is not a standard practice and there is not yet a commonly agreed procedure for models evaluation and comparison. Recently Mak et al. (2014) pointed out that, depending on the limited time span of available observations (compared with the selected return period of PSHA map), the probability of failing to reject an inadequate model can be high. Even if formal testing does not guarantee the adequacy of a model, a quantitative analysis of performances may allow comparing different models and spotting out possible problems. Objective testing, in fact, may have different purposes, ranging from purely scientific verification of model distributions and parameters to the assessment of maps predictive capability for moderate to extreme shaking, which may require specific metrics and tests.

In spite of the evidenced shortcomings and of its poor performances (see Panza et al., 2014 for an in depth discussion), PSHA is still widely applied in the framework of several large scale projects at regional and global scale (e.g. Global Earthquake Model). Most of such attempts in improving seismic hazard maps, however, basically rely on the collection and revision of the input data and, so far, did not include a formal procedure to assess the improved capability of the revised maps in describing ground shaking. By analogy with medicine testing, in fact, the adequacy of the proposed maps should be established before their publication and control should be performed by the proponents as a primary test of reliability of the new results.

A possible alternative to the conventional PSHA approach is provided by the Neo-Deterministic Seismic Hazard Assessment, NDSHA (Panza et al. 2001; 2012; 2013), a methodology that allows for the consideration of a wide range of possible seismic sources as the starting point for deriving scenarios via full waveforms modeling. Besides the standard NDSHA maps, which provide reliable estimates of maximum seismic ground motion from a wide set of possible scenario earthquakes, the flexibility of NDSHA allows to account for earthquake recurrence and it permits to compute ground shaking maps at specified return periods (Peresan et al., 2013). A systematic comparative analysis was carried out for the territory of Italy between the NDSHA and PSHA maps (the last is at the base of current seismic regulation), investigating their performances with respect to past earthquakes, so as to better understand the performances and possible limits of the two different approaches to seismic hazard assessment (Nekrasova et al., 2014).

In this study the comparative analysis is extended to additional hazard maps for the Italian territory, which are available from large scale projects (i.e. GSHAP), including the most recent probabilistic map, which has been compiled for the territory of Europe in the framework of Seismic Hazard Harmonization in Europe (SHARE) project. The new European Seismic Hazard Map (ESHM13), in fact, has been released recently by Giardini et al. (2014) with the following declared intent:

> "SHARE's main objective is to provide a community-based seismic hazard model for the Euro-Mediterranean region with update mechanisms. The project aims to establish new standards in Probabilistic Seismic Hazard Assessment (PSHA) practice by a close cooperation of leading European geologists, seismologists and engineers."

Regretfully, the new SHARE map does not seem to address most of the limits of the PSHA approach (e.g. Stein et al., 2012, Panza et al., 2014) and repeats the errors of its predecessors, possibly (mis)leading to unexpected economic and human life losses from future earthquakes.

## 2. Data

In this study we consider ground shaking estimates for the territory of Italy within the boundaries from 36°N to 48°N and from 6°E to 20°E provided by the following eight seismic hazard assessment maps.

(a) The final Global Seismic Hazard Assessment Program (GSHAP) map that depicts peak ground acceleration (PGA) values with a 10% chance of exceedance of in 50 years (GSHAP10%) corresponding to a return period of 475 years.

The GSHAP PGA values obtained by the probabilistic seismic hazard analysis (PSHA) methodology and presented as the final Global Seismic Hazard Map (Shedlock et al., 2000; Giardini et al., 2003) and Table (GSHPUB.dat, http://www.seismo2009.ethz.ch/GSHAP/) are provided on a 0.1°×0.1° regular grid for seismically active regions of the Globe, including the territory of Italy.

(b) The SHARE PGA values as defined by a 10% chance of exceedance in 50 years (SHARE10%) corresponding to a return period of 475 years.

(c) The SHARE PGA values for a probability of exceedance of 2% in 50 years (SHARE2%) associated with a 2475-year return period.

The SHARE PGA values, obtained by the, claimed, improved PSHA methodology, are given at the grid points of a regular 0.1°×0.1° mesh, which data can be downloaded from http://www.efehr.org:8080/jetspeed/portal/hazard.psml.

(d) The current Italian official seismic hazard map PGA values as defined by a 10% chance of exceedance of in 50 years (PGA10%) corresponding to a return period of 475 years

(e) The Italian official seismic hazard map PGA values for a 2% probability of exceedance in 50 years (PGA2%) associated with a return period of 2475 years.

Both the official seismic hazard maps for Italy are based on PSHA (Meletti and Montaldo, 2007; the data file http://esse1.mi.ingv.it/d2.html) at the grid points of a regular 0.2°×0.2° mesh.

(f) The maximum design ground acceleration (DGA) map for Italy, estimated by the standard NDSHA approach.

(g) The NDSHA DGA values estimated for a return period of 475 years, corresponding to a 10% chance of exceedance of in 50 years (DGA10%).

(h) The NDSHA DGA values estimated for a return period of 2475 years, corresponding to a 2% chance of exceedance of in 50 years (DGA2%).

The three design ground acceleration (DGA) maps are based on the neo-deterministic seismic hazard assessment, NDSHA (Panza et al., 2012 and references therein), which provides ground shaking estimates at the grid points of a regular 0.2°×0.2° mesh. From the complete synthetic seismograms associated to each grid point, the DGA estimates are extracted, which can be compared to PGA (Zuccolo et al., 2011). The DGA map defined by the standard NDSHA method does not depend on temporal properties of earthquakes occurrence, whereas the DGA10% and DGA2% maps are obtained by incorporating earthquake recurrence information into NDSHA (Peresan et al., 2013; Magrin, 2012), and correspond to return periods of 475 and 2475 years, respectively (i.e. same as considered in compilation of the PSHA maps). The application of NDSHA variant that computes ground shaking at a fixed return period implies additional requirements to the input data, which are not fulfilled in the parts of the Italian territory delineated as blank areas in Fig.1 g, h. In turn, the limits of available data in adequately constraining ground shaking recurrence, as evidenced by NDSHA analysis (Peresan et al. 2013), casts doubts on the meaning and validity of PSHA values given for these blank areas, if based on the same data.

For the purpose of comparison between different grids we enhance the regular 0.2°×0.2° mesh into a 0.1°×0.1° one, so that each PGA value from the original grid point is attributed to four points on the fine grid (i.e. the original point, plus its three nearest neighbors to the east, south, and south-east).

The observed seismic activity data are taken from the SHARE European Earthquake Catalogue (SHEEC), as reported by Stucchi et al. (2012) for historical events in 1000-1899 and by Grünthal et al. (2014) for earthquakes in 1900-2006. The data set covering more than a millennium (a time interval about ten times longer than that available in most of the regions worldwide), with a completeness level satisfactory for this kind of analysis, is quite a unique property of the territory of Italy and fully warrants the following analysis. The SHEEC data provides records on macroseismic intensity at epicenter, $I_0$. In our analysis we have used integer values of $I_0$, attributing the upper limit when in SHEEC the reported $I_0$ is a range. This is a conservative natural choice of seismic hazard estimate, adequate to analysis aimed at the largest possible ground shaking. The observed intensity map, $I_{obs}$, is compiled by attributing to a grid point of a regular 0.1°×0.1° mesh the maximum of $I_0$ for earthquakes from SHEEC within the 0.25°-side square centered at this grid point. This resulting map of the observed ground shaking intensity gives us an opportunity for a quantitative comparison of the eight seismic hazard maps of the Italian territory with the seismic reality.

### 3. Cross-comparison of the PGA maps for Italy.

We repeat the analysis reported in (Nekrasova et al., 2014), expanding the comparison to the probabilistic seismic hazard maps for the Italian territory obtained in the framework of large scale projects: the Global Seismic Hazard Assessment Programme (GSHAP) map published 15 years ago (Giardini et al., 1999), and its new offspring for Europe (SHARE), which became available recently (Giardini et al., 2014).

Table 1 gives an overall summary on the PGA values for each of the eight SHA maps (Fig. 1). Evidently, the SHARE maps increase dramatically both the lower and the upper limits of the expected seismic hazard in Italy. In particular, the minimum of the SHARE PGA is about 2 and 4 times larger than the corresponding estimates of the earlier probabilistic SHA. In comparison to the NDSHA maps the minimum values of ground shaking by the SHARE maps are about 5 and 10 times larger. The increase of the maximum PGA on the SHARE maps accounts to about 10-50% of the corresponding previously suggested values.

**Table 1** The parameters of the eight SHA ground acceleration maps for Italian territory.

| Map | GSHAP | SHARE10% | SHARE2% | PGA10% | PGA2% | DGA | DGA10% | DGA2% |
|---|---|---|---|---|---|---|---|---|
| Number of points | 3066 | 3066 | 3066 | 3044 | 3044 | 3066 | 1739 | 2266 |
| min(mGA), $m/s^2$ | 0.39 | 0.74 | 1.72 | 0.30 | 0.43 | 0.20 | 0.16 | 0.18 |
| max(mGA), $m/s^2$ | 2.97 | 4.19 | 8.82 | 2.71 | 5.98 | 5.83 | 3.74 | 5.83 |

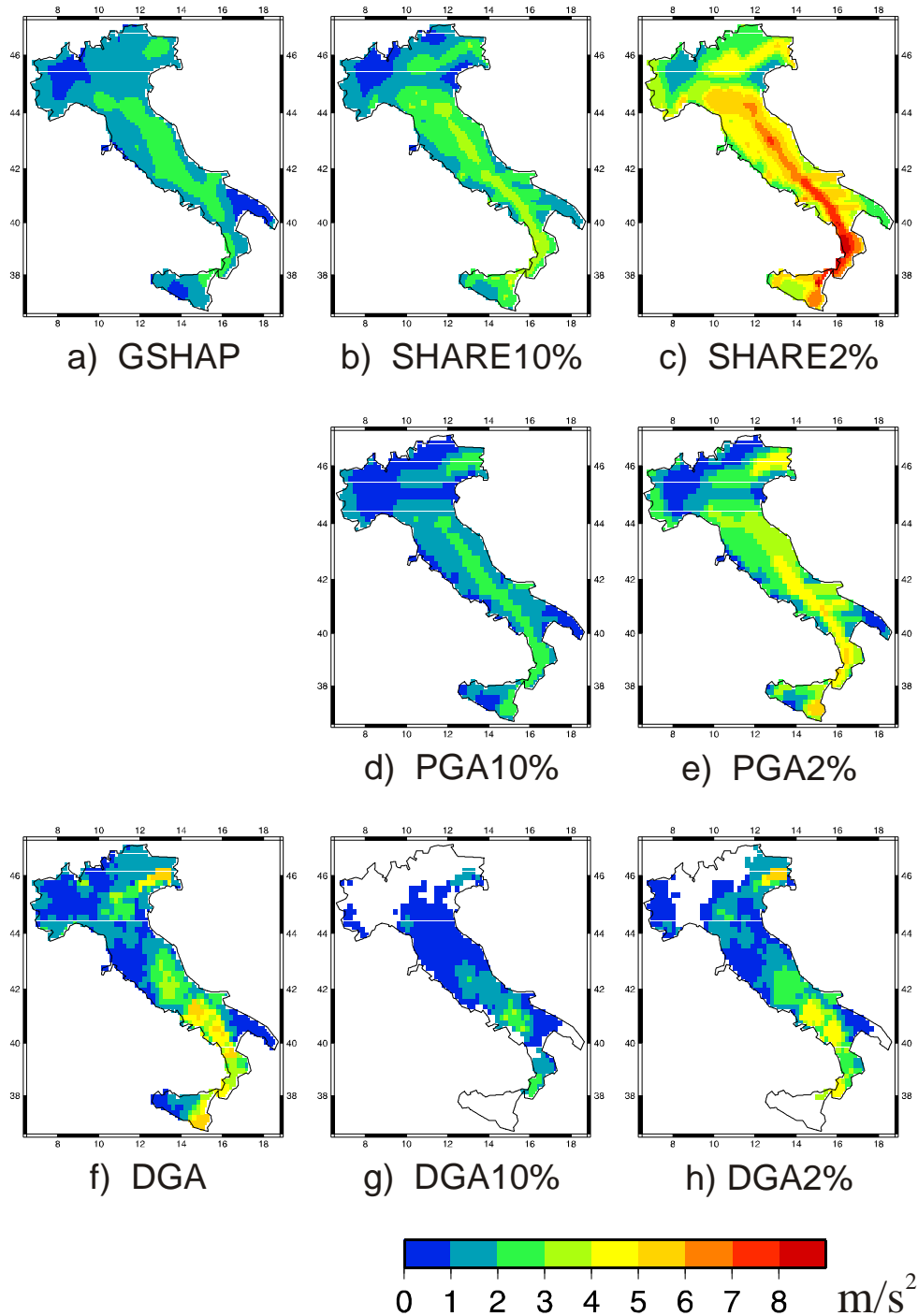

**Fig. 1** Comparison of model ground acceleration maps: a) GSHAP corresponding to a return period of 475 years; b) SHARE10% corresponding to a return period of 475 years; c) SHARE2% corresponding to a return period of 2475 years; d) PGA10% corresponding to a return period of 475 years; e) PGA2% corresponding to a return period of 2475 years; f) DGA not depending on time; g) DGA10% corresponding to a return period of 475 years; h) DGA2% corresponding to a return period of 2475 years.

Table 2 provides a more refined comparison, based upon the percentage of PGA values ratio at a grid point (mGA1/mGA2), for a number of pairs of model maps mGA1 and mGA2. SHARE map values at a grid point exceed those of the previous maps by a factor of 2 or more in 4% of cases, for the GSHAP map, in 17% and 40% of cases, for the corresponding national PSHA maps, and to up to more than 75% of cases, for the NDSHA estimates. The SHARE estimate is less than any previous hazard estimates in 2%-3% of grid points. Specifically, as can be concluded from the maps of the ratio of the PGA values for different pairs of models, a selection being provided in Figure 2, just about 2% of the grid points of the Italian official SHA maps have higher PGA values than that of SHARE; these are all located in the Friuli-Venezia-Giulia region (Figure 2 c-d). In comparison to the previous hazard maps, the SHARE PGA values corresponding to a return period of 475 years increase by a factor of 2 or more in the regions of Trentino, Lombardia, Eastern Sicily, and Puglia; for Liguria PGA increases more than 4 times. The misfit of the SHARE maps with respect to the NDSHA ones is even more dramatic (Figure 2 e-g): e.g. the SHARE2% values are larger than the DGA2% by a factor of 4 or more in about 40% of the Italian territory.

**Table 2** The percentage of the grid points from different ranges of the ratio mGA1/mGA2 of the PGA values from selected pairs of SHA maps.

| $\dfrac{mGA1}{mGA2}$ range | GSHAP / PGA10% | SHARE10% / GSHAP | SHARE10% / PGA10% | SHARE2% / PGA2% | SHARE10% / DGA | SHARE10% / DGA10% | SHARE2% / DGA2% |
|---|---|---|---|---|---|---|---|
| ≥4 | - | - | 0.26 | 1.25 | 4.04 | 30.82 | 39.81 |
| ≥2 | 6.73 | 4.21 | 17.05 | 40.31 | 26.65 | 75.79 | 78.60 |
| ≥1 | 78.58 | 85.32 | 97.40 | 97.96 | 65.04 | 97.99 | 97.04 |
| <1 | 21.42 | 14.68 | 2.60 | 2.04 | 34.96 | 2.01 | 2.96 |

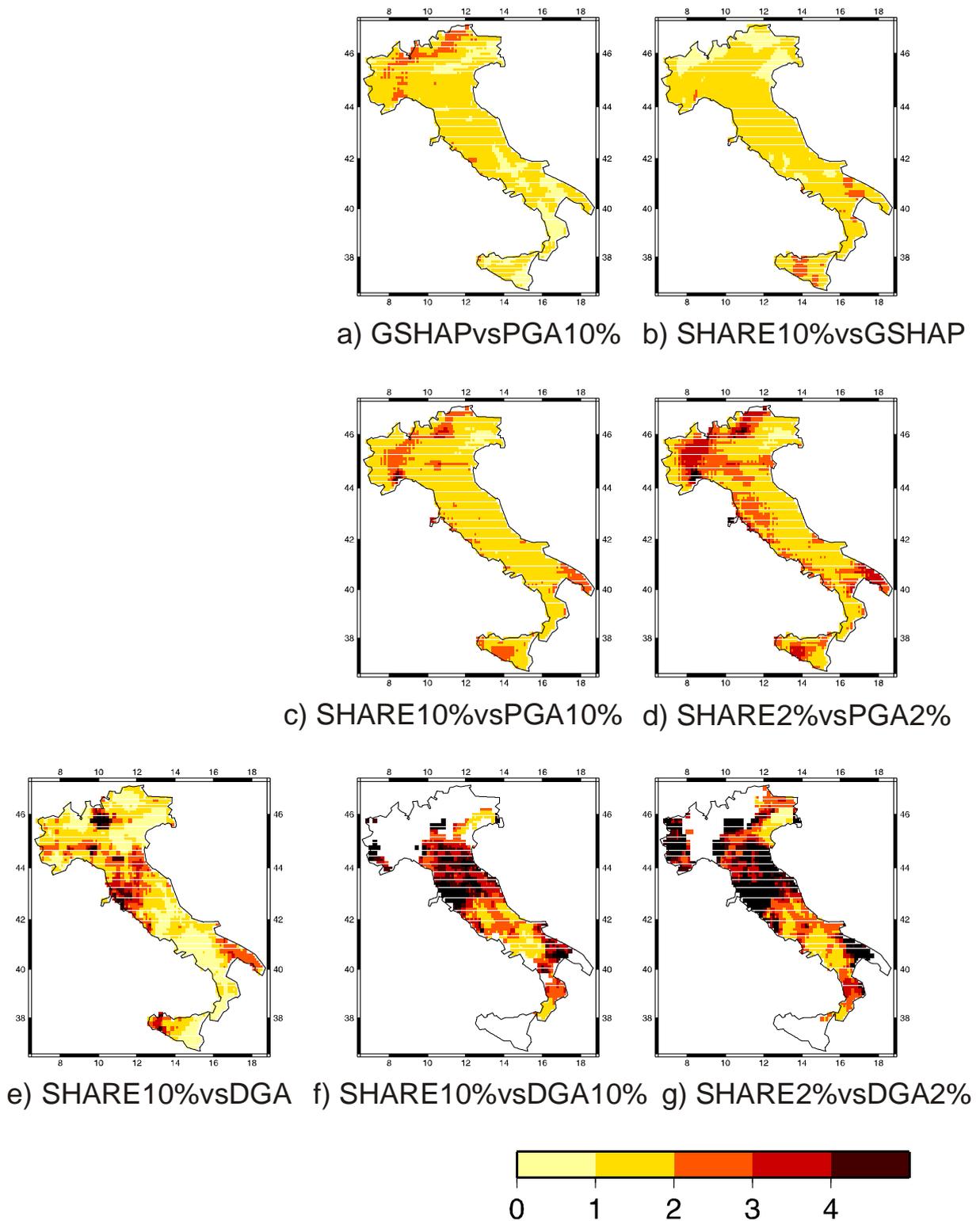

**Fig. 2** Selected maps of the ratio between PGA values from different pairs of SHA maps: a) GSHAP divided by PGA10%; b) SHARE10% divided by GSHAP; c) SHARE10% divided by PGA10%; d) SHARE2% divided by PGA2%, e) SHARE10% divided by DGA f) SHARE10% divided by DGA10%, g) SHARE2% divided by DGA2%.

Each of the six graphs in Figure 3 shows the correlation diagram between a pair of seismic hazard maps of Italy, displaying the PGA values on a grid point of one map versus the PGA values on the same grid point of another map. These are all possible pairs of maps corresponding to a return period of 475 years (i.e. probabilistic GSHAP, SHARE10%, PGA10%, and neo-deterministic DGA10%). It is evident that the most recent map reviewed by the probabilistic approach to seismic hazard assessment (SHARE10%) evidently provides a gross overestimation of PGA values, compared to all of the other maps. The neo-deterministic map (DGA10%) is the most optimistic in providing low PGA values, under 1 m/s$^2$, but conservative in expecting high accelerations, above 2 m/s$^2$.

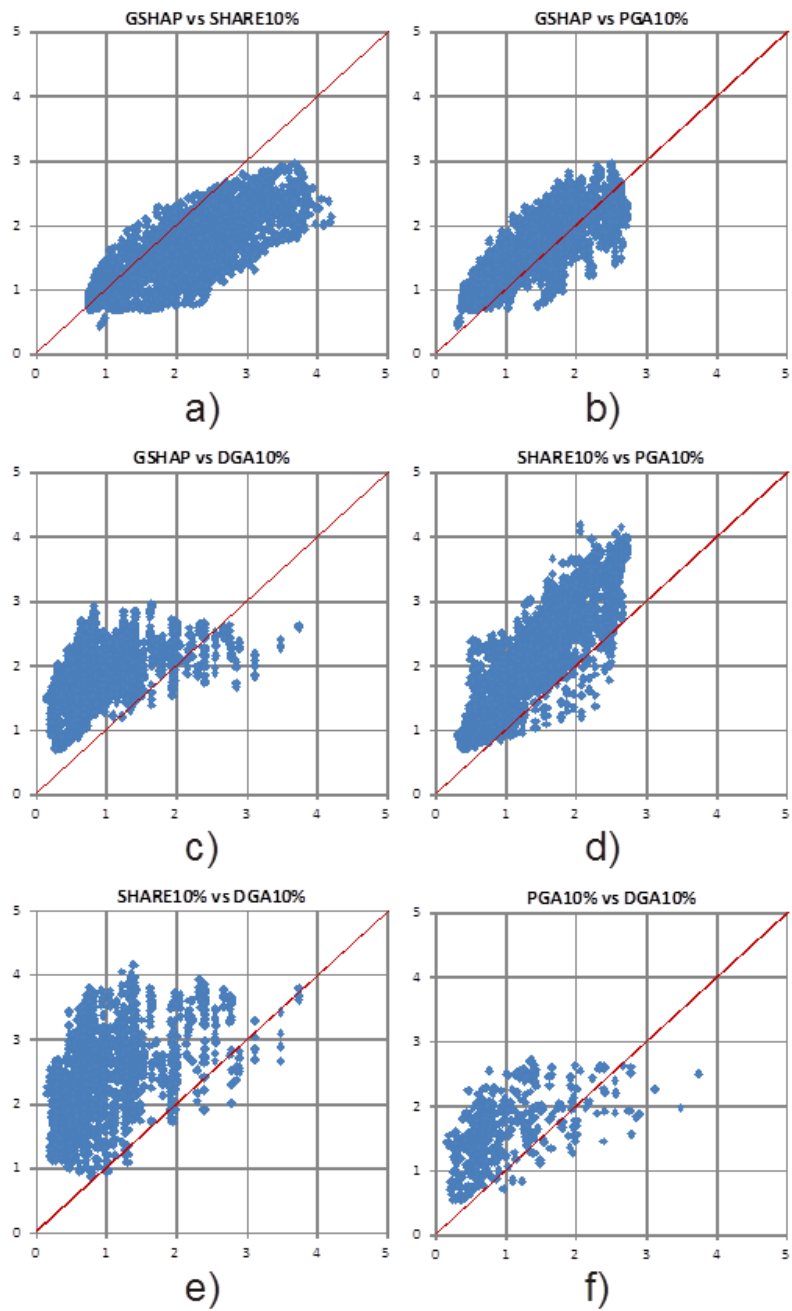

**Fig. 3** Correlation diagrams of the PGA values (in m/s$^2$) on the four hazard maps of Italy corresponding to a return period of 475 years: a) GSHAP (ordinate) versus SHARE10% (abscissa); b) GSHAP versus PGA10%; c) GSHAP versus DGA10%; d) SHARE10% versus PGA10%; e) SHARE10% versus DGA10%; f) PGA10% versus DGA10%. These are all possible pairs of maps corresponding to a return period of 475 years (i.e. probabilistic GSHAP, SHARE10%, PGA10%, and adjusted neo-deterministic DGA10%).

Of course, any cross-comparison of the maps obtained by different models and/or series of correlation diagrams does not answer to the key question of a model adequacy to reality. In the next section we address this pivotal question in hazard assessment by comparisons of the model maps with the available observations.

## 4. Comparison of the hazard maps for Italy against registered ground shaking.

The two currently official seismic hazard maps for Italy PGA10% and PGA2% and the three neo-deterministic maps DGA, DGA10% and DGA2% were already subject of comparison in (Nekrasova et al, 2014). Here we update and expand the comparison with the observed ground shaking to the European Seismic Hazard Maps 2013 - SHARE10% and SHARE2%, issued recently (Giardini et al. 2014), along with their predecessor, GSHAP map (Giardini et al. 1999). Table 3 lists the conversion rules between PGA and MCS for the territory of Italy after Indirli et al. (2011). These rules are used to convert the estimated ground shaking from SHARE10%, SHARE2%, PGA10%, PGA2%, DGA, DGA10%, DGA2% into the corresponding macroseismic intensity MCS values. Figure 4 shows the eight model intensity maps subject to comparison along with the map $I_{obs}$ compiled from the SHEEC reported data. All the nine intensity maps refer to the same regular 0.1°×0.1° mesh within the borders of Italy. For the purposes of comparison the recurrence adjusted DGA10% and DGA2% neo-deterministic maps were expanded to the grid points of no recurrence determination, following the empirical linear regression equation that links the DGA map values and the existing estimates on the DGA2% and DGA10% maps. The resulting model intensity maps are DGA2%* and DGA10%*, respectively.

Figure 4 presents the eight intensity maps obtained (i) from the real seismicity $I_{obs}$ (Figure 4a) as well as (ii) from the ground motion estimates $I_{SHARE10\%}$, $I_{SHARE2\%}$, $I_{PGA10\%}$, $I_{PGA2\%}$, $I_{DGA}$, $I_{DGA10\%*}$, $I_{DGA2\%*}$ (Figure 4b-h, respectively).

**Table 3** Relation between $I_{MCS}$ and model ground motion, mGA, after Indirli et al. (2011)

| $I_{MCS}$ | VI | VII | VIII | IX | X | XI |
|---|---|---|---|---|---|---|
| mGA, (g) | 0.01-0.02 | 0.02-0.04 | 0.04 – 0.08 | 0.08 – 0.15 | 0.15-0.3 | 0.3-0.6 |

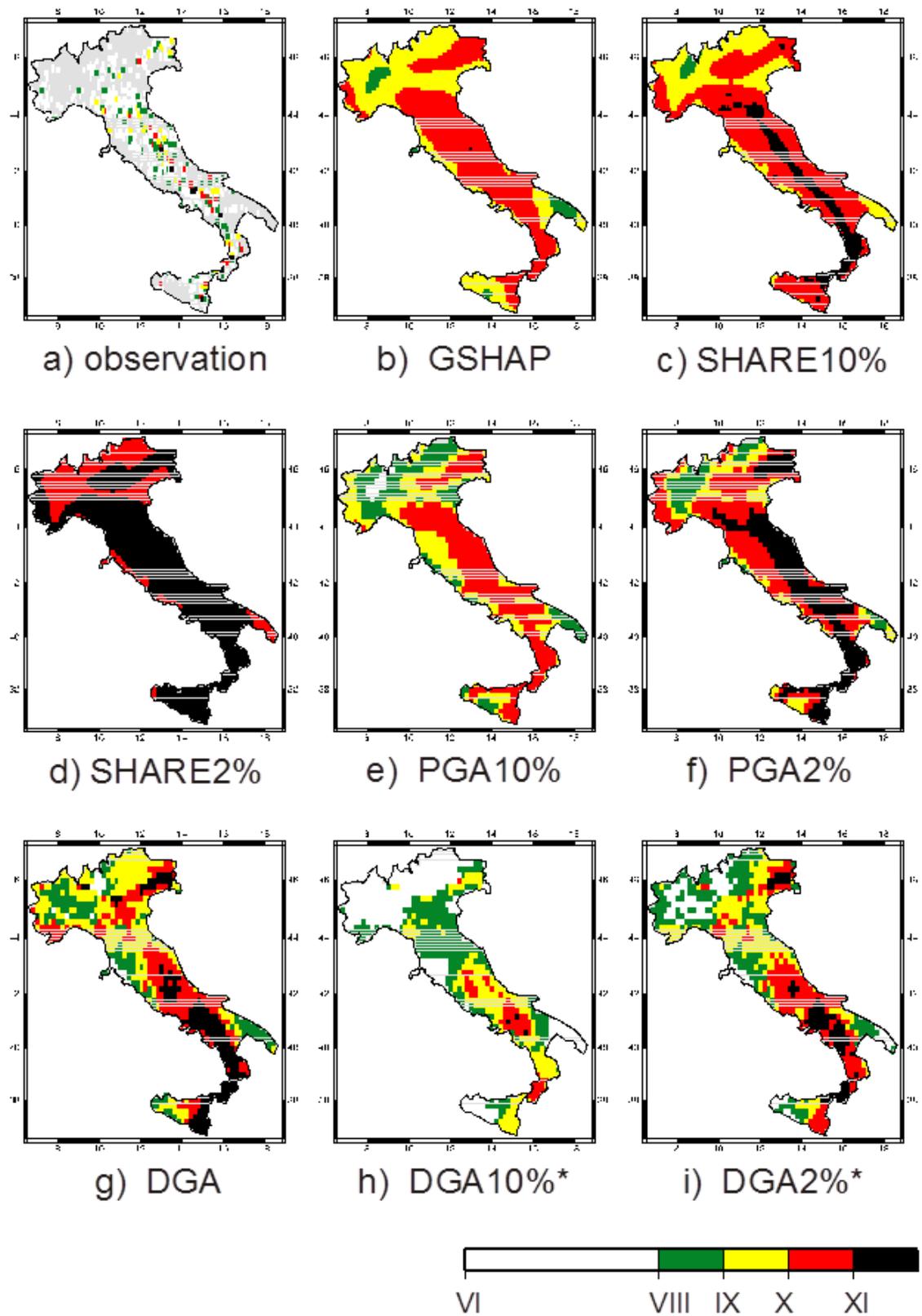

**Fig.4** The intensity maps under comparison: a) $I_{obs}$ map obtained from the reported seismicity data in 1000-2006; model intensity corresponding to PGA maps – b) GSHAP, c) SHARE10%, d) SHARE2%, e) PGA10%, f) PGA2%, g) DGA, h) DGA10%*, and i) DGA2%*.

**Table 3** The percentage of $I_{MCS}$ from different ranges in real observation map ($I_{obs}$) and in the model intensity maps corresponding to the eight hazard maps considered.

| $I_{MCS}$ range | $I_{obs}$ | GSHAP | SHARE10% | SHARE2% | PGA10% | PGA2% | DGA | DGA10%* | DGA2%* |
|---|---|---|---|---|---|---|---|---|---|
| ≥ XI | 2.76 | 0.03 | 14.58 | 75.28 | - | 33.84 | 19.28 | 0.39 | 10.21 |
| ≥ X | 11.63 | 56.20 | 72.70 | 100 | 44.48 | 74.15 | 45.66 | 8.06 | 34.93 |
| ≥ IX | 21.40 | 94.98 | 98.56 | 100 | 76.74 | 90.57 | 78.90 | 31.12 | 59.52 |
| ≥ VIII | 41.69 | 100 | 100 | 100 | 97.77 | 100 | 97.23 | 61.42 | 86.82 |
| ≥ VII | 74.27 | 100 | 100 | 100 | 100 | 100 | 100 | 86.43 | 98.47 |
| ≥ VI | | | | | 100 | | | | |

The percentage of the points with intensity VI or more for each of these maps is summarized in Table 3. Remarkably, the SHARE2% assigns all the territory to "extreme" ground shaking of intensity X or larger, while the $I_{obs}$ map of macroseismic intensities reports such intensity, in about two thousand years of observations, for less than 12% of the territory. At this "extreme" level of ground shaking the DGA10%* with its 8% appear to be the nearest to $I_{obs}$, and, in general, the neo-deterministic maps are closer to reality than all the probabilistic ones but PGA10%, which predicts (about 45% of intensity X) for a return period of 475 years similar values to those of the time unlimited DGA. Similar situation exists at the "severe", intensity VIII level of ground shaking: it is attributed to 100% of the Italian territory by all the probabilistic maps except PGA10%, which attributes it to 98% of the territory, still too large in comparison to 42% of $I_{obs}$. Once again the DGA10%* with its 61.42% is the closest to $I_{obs}$.

More rigid comparison with respect to the $I_{obs}$ map can be performed by applying the Kolmogorov-Smirnov test that quantifies the distance between the empirical distribution functions. The maximum absolute difference between the empirical distributions is commonly used in the Kolmogorov-Smirnov two-sample criterion to distinguish whether or not the values from the two samples are drawn from the same statistical distribution of independent variables. We apply the two sample Kolmogorov-Smirnov statistic $\lambda_{K-S}$ to the empirical distribution functions of MCS values on a model map and the observed SHEEC reported data map:

$\lambda_{K-S}(D, n, m) = [nm/(n+m)]^{1/2} D$,

where $D = \max | F_i(I) - F_0(I) |$ is the maximum of the absolute difference between the empirical distributions of the i-th model map $F_i(I)$ and the $I_{obs}$ map $F_0(I)$, whose sample sizes are *n* and *m*, respectively; *I* = VI, VII, VIII, IX, X, XI, XII. Figure 5a shows the empirical distribution functions used in the comparison. For the purposes of additional testing and qualitative uncertainty estimation, the empirical distribution function of the MCS values from the publicly available database of direct macroseismic observations DBMI04 (Stucchi et al. 2007) is also object of comparison with the $I_{obs}$ map. Figure 5b shows the nine differences $F_i(I) - F_0(I)$ and it illustrates the departure of a model from the zero-line of reality; the departure of direct MCS observations from DBMI04 characterizes the realistic dispersion of the real data. Table 4 summarizes the results of the comparison in terms of computed $D$ and $\lambda_{K-S}$.

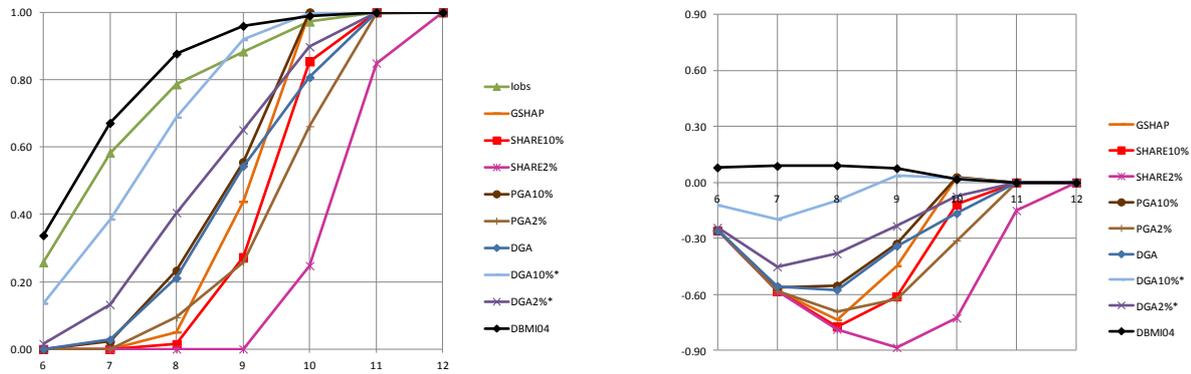

**Fig. 5** The empirical probability functions of macroseismic intensity (a) and the difference between a model and the real intensities $F_i(I) - F_0(I)$ (b).

**Table 4** The Kolmogorov-Smirnov two-sample statistic $\lambda_{K-S}$ applied to a model map and the real seismic intensity map ($I_{obs}$, sample size 1341). Sample size indicates the number of grid points analysed.

| Statistic | Model seismic intensity map | | | | | | | | |
|---|---|---|---|---|---|---|---|---|---|
| | GSHAP | SHARE10% | SHAE2% | PGA10% | PGA2% | DGA | DGA10%* | DGA2%* | DBMI04 |
| Sample size | 3066 | 3066 | 3066 | 3044 | 3044 | 3066 | 3066 | 3066 | 19713 |
| $D$ | 0.74 | 0.77 | 0.88 | 0.56 | 0.69 | 0.57 | 0.20 | 0.45 | 0.09 |
| $\lambda_{K-S}$ | 22.47 | 23.57 | 26.99 | 17.11 | 21.10 | 17.56 | 6.03 | 13.79 | 1.82 |

The K-S test results confirm quantitatively the conclusions that could have been already reached from Table 3: the values of seismic intensity assigned by any of the models considered and reported in SHEEC do not come from the same distribution. The $\lambda_{K-S}$ for the two representatives of the observed ground shaking, i.e. at epicenter (the $I_{obs}$ map) and at site of direct observation (the DBMI04 data), provides an empirical estimate of admissible departure of real intensity distributions as a reference for claiming consistency of a model. Nekrasova et al (2014) have shown that the DGA10% map appears to be "the best fit" among the five model intensity maps for Italy (i.e., the two official and the three neo-deterministic maps). The investigation expanded to the eight model maps confirms this conclusion. Moreover, it becomes evident that the GSHAP and most recent SHARE maps for Italy are hardly consistent with observations and overestimate dramatically the seismic hazard in the region. Apparently the SHARE maps keep moving away from reality, even more than GSHAP.

Tables 5 and 6 disclose the quality of a model map in predicting the maximum of the macroseismic intensity, in particular, the expectation of "a 10% or 2% chance of exceedance in 50 years". Table 5 indicates clearly that for SHARE maps the number of exceedances, by a unit of MCS intensity or larger, is by far smaller than one should expect from the number of trials represented by the intensity VIII or larger records in DBMI04. Once again the fit of the adjusted neo-deterministic DGA10%* and DGA2%* (11.6% and 1.9% exceedances, correspondingly) is more consistent with expectations than that of the probabilistic maps, both for a return period of 475 years (GSHAP is exceeded in 4.4%, SHARE10% in 0.1% and PGA10% in 4.5% of cases, respectively) and for a return period of 2475 years (SHARE2% and PGA2% are never exceeded, thus it is 0.0%). The small sample of $I_0 \geq$ VIII earthquakes from SHEEC (Table 6) does not permit, although does not contradict, the conclusion on a model map consistency, as clear as with the macroseismic records from DBMI04.

**Table 5**. Number of grid points where the difference Δ, between the intensity $I \geq$ VIII records in DBMI04 and the maximum of the model map values at the distance of 1/8° or less, has been computed.

| Model | GSHAP | SHARE10% | SHARE2% | PGA10% | PGA2% | DGA | DGA10%* | DGA2%* |
|---|---|---|---|---|---|---|---|---|
| Total | 4421 | 4421 | 4421 | 4421 | 4421 | 4421 | 4421 | 4421 |
| Δ = 2 | 1 | 0 | 0 | 0 | 0 | 0 | 81 | 6 |
| Δ = 1 | 193 | 3 | 0 | 200 | 0 | 9 | 434 | 76 |
| Δ = 0 | 589 | 236 | 0 | 620 | 189 | 276 | 1194 | 430 |
| Δ = -1 | 1562 | 706 | 257 | 1611 | 633 | 963 | 1444 | 1125 |
| Δ = -2 | 2009 | 1790 | 914 | 1990 | 1670 | 1616 | 1049 | 1557 |
| Δ = -3 | 67 | 1686 | 1711 | 0 | 1924 | 1557 | 213 | 1227 |
| Δ = -4 | 0 | 0 | 1539 | 0 | 5 | 0 | 0 | 0 |

**Table 6**. Number of earthquakes with $I_0 \geq$ VIII from SHEEC for which the difference Δ, between $I_0$ and the maximum of the model map values at the distance of 1/8° or less, has been computed.

| Model | GSHAP | SHARE10% | SHARE2% | PGA10% | PGA2% | DGA | DGA10%* | DGA2%* |
|---|---|---|---|---|---|---|---|---|
| Total | 42 | 42 | 42 | 42 | 42 | 42 | 42 | 42 |
| Δ = 2 | 0 | 0 | 0 | 0 | 0 | 0 | 3 | 1 |
| Δ = 1 | 2 | 0 | 0 | 1 | 0 | 1 | 5 | 1 |
| Δ = 0 | 8 | 2 | 0 | 9 | 1 | 5 | 8 | 5 |
| Δ = -1 | 12 | 10 | 4 | 13 | 9 | 9 | 19 | 12 |
| Δ = -2 | 20 | 16 | 9 | 19 | 16 | 13 | 6 | 14 |
| Δ = -3 | 0 | 14 | 18 | 0 | 16 | 14 | 1 | 9 |
| Δ = -4 | 0 | 0 | 11 | 0 | 0 | 0 | 0 | 0 |

## 5. Conclusions

The comparison of the model intensity maps against the real seismic activity in Italy, made over a time interval of more than a millennium, reveals many discrepancies in several aspects of the models seismic ground shaking distribution in space and size.

We did repeat the analysis reported in (Nekrasova et al., 2014) and expanded it to the Global Seismic Hazard Assessment Programme (GSHAP) map and its new offspring for Europe (SHARE), which became available recently (Giardini et al., 2014). The results of the analysis described in this paper confirm the following conclusions:

- the estimates of seismic intensity attributed by any of the eight models considered, including the official and most recent SHARE seismic hazard maps, and those reported in the Italian databases of empirical observations could hardly arise from the same distribution;
- models (except for the recurrence adjusted neo-deterministic DGA10% and DGA2%, at the cost of model consistent 10% or 2% cases of exceedance) generally provide rather conservative estimates with respect to reality. They tend to over-estimate seismic hazard particularly at the levels below violent (MCS intensity IX) ground shaking events and yet most of them do not guarantee avoiding underestimations for the largest earthquakes;
- probabilistic maps have a higher tendency to overestimate the hazard, with respect to the corresponding deterministic maps and reality; in particular, the newly published SHARE maps assign practically all the territory of Italy to extreme ground shaking of intensity $I \geq X$;
- in terms of the goodness of fit measured by the Kolmogorov-Smirnov two-sample statistic, the NDSHA models appear to outscore the probabilistic ones and might be a better representation of the real seismicity. In particular, the minimum value of $\lambda_{K\text{-}S}$ obtained for DGA10%* is 3-4 times smaller than for any of the probabilistic models, while it is 3 times larger than for the reference misfit of the observed ground shaking at epicenters and at sites of direct observations.

The study of the statistical significance of the detected inconsistencies between model and observed intensities and their interpretation should be addressed in further investigation of earthquake phenomenon, in particular for the predictability of the maximum ground shaking.

What is often the problem with probabilistic approaches is that probability is a purely mathematical concept and, by the law of large numbers, the frequency approaches the probability only in an infinite collection of independent identically distributed random occurrences. It is clear, therefore, that all methods that mix these two terms (frequency and probability) without computing the deviations with sufficient number of moments are bound to fail sooner or later. As expected, an oversimplified model computation of the minimum time interval required for reliable occurrence rate estimates with reasonable uncertainty for a return period of 475 years (Beauval et al. 2008; Mak et al. 2014) suggests the geological time span of 12,000 years. In the case of Italy, on account of the millennial earthquake catalogue available, a reliable and physically sound alternative is represented by NDSHA hazard estimations, which in their standard definition do not depend on the probability of earthquake occurrence, but can be adjusted by recurrence if the data allow.

The obtained results might be indicative of a fundamental misfit of the generally accepted uniform rules of homogeneous smoothing applied to observations on top the naturally fractal system of blocks-and-faults with evidently heterogeneous structure and rheology. Any model for SHA aimed, presumably, at predicting disastrous ground shaking that would actually occur must pass series of rigid testing against the available real seismic activity data before being suggested as a practical seismic hazard and risk estimation. Otherwise, similar to medical malpractice, although at much higher level of simultaneous losses (Wyss et al. 2012), the use of untested seismic hazard maps would eventually mislead to crime of negligence.

**Acknowledgements**

This paper was completed during the visit of A.K. Nekrasova at the Structure and Nonlinear Dynamics of the Earth (SAND) Group of the Abdus Salam International Centre for Theoretical Physics, Miramare - Trieste, Italy. AKN and VGK acknowledge the support from the Russian Foundation for Basic Research (RFBR grants № 13-05-91167 and № 14-05-92691).

**References**

Beauval C., P-Y. Bard P-Y, S. Hainzl, Guguen (2008) Can strong motion observations be used to constrain probabilistic seismic hazard estimates? Bull Seismol Soc Am 98(2): 509-520.

Giardini D., G. Grünthal, K.M. Shedlock, P. Zhang (1999) The GSHAP Global Seismic Hazard Map. Annali di Geofisica 42 (6): 1225-1228.

Giardini D., G. Grünthal, K.M. Shedlock, P. Zhang (2003) The GSHAP Global Seismic Hazard Map. In: Lee, W., Kanamori, H., Jennings, P. and Kisslinger, C. (eds.): International Handbook of Earthquake & Engineering Seismology, International Geophysics Series 81 B, Academic Press, Amsterdam, 1233-1239.

Giardini D., J. Woessner, L. Danciu, F. Cotton, H. Crowley, G. Grünthal, R. Pinho, G. Valensise, S. Akkar, R. Arvidsson, R. Basili, T. Cameelbeck, A. Campos-Costa, J. Douglas, M. B. Demircioglu, M. Erdik, J. Fonseca, B. Glavatovic, C. Lindholm, K. Makropoulos, C. Meletti, R. Musson, K. Pitilakis, A. Rovida, K. Sesetyan, D. Stromeyer, M. Stucchi, (2013) Seismic Hazard Harmonization in Europe (SHARE): Online Data Resource, doi:10.12686/SED-00000001-SHARE

Giardini D., J. Woessner, L. Danciu (2014) Mapping Europe's Seismic Hazard, Eos Trans. AGU, Eos, Vol. 95, No. 29, 22 July 2014

Grünthal G., R. Wahlström, D. Stromeyer (2013) The SHARE European Earthquake Catalogue (SHEEC) for the time period 1900-2006 and its comparison to the European Mediterranean Earthquake Catalogue (EMEC). Journal of Seismology (submitted).

Indirli M, H. Razafindrakoto, F. Romanelli, C. Puglisi, L. Lanzoni, E. Milani, M. Munari, S. Apablaza (2011) Hazard Evaluation in Valparaiso: the MAR VASTO Project. Pure Appl Geophys 168(3-4):543-582

Kossobokov V.G., A.K. Nekrasova (2012). Global seismic hazard assessment program maps are erroneous. Seismic Instrum., 48 (2). http://dx.doi.org/10.3103/S0747923912020065. Allerton Press, Inc., 2012162-170.

Magrin A. (2012) Multi-scale seismic hazard scenarios., PhD Thesis. Univ. degli Studi di Trieste, Italy.

Mak S., R.A. Clements, D. Schorlemmer (2014) The statistical power of testing probabilistic seismic-hazard assessments. Seismological Research Letters 85(4): 781-783.

Marzocchi W. (2013), Seismic Hazard and Public Safety. Eos, Vol. 94, No. 27, 240-241.

Meletti C., V. Montaldo (2007) Stime di pericolosità sismica per diverse probabilità di superamento in 50 anni: valori di ag. http://esse1.mi.ingv.it/d2.html, Deliverable D2

Nekrasova A., V. Kossobokov, A. Peresan, A. Magrin (2014) The comparison of the NDSHA, PSHA seismic hazard maps and real seismicity for the Italian territory, Natural Hazards 70 (1), pp.629–641 DOI 10.1007/s11069-013-0832

Panza G.F., F. Romanelli, F. Vaccari (2001) Seismic wave propagation in laterally heterogeneous anelastic media: Theory and applications to seismic zonation. Advances in Geophysics, 43:1-95.

Panza G.F, La Mura C, Peresan A, Romanelli F, Vaccari F. (2012). Seismic Hazard Scenarios as Preventive Tools for a Disaster Resilient Society. In: Dmowska R (Ed) Advances in Geophysics, Elsevier, London, 93–165.

Panza G.F., A. Peresan A., C. La Mura (2013). Seismic Hazard and Strong Ground Motion: an Operational Neo-deterministic Approach from National to Local Scale. [Eds.UNESCO-EOLSS Joint Commitee], in Encyclopedia of Life Support Systems (EOLSS), Geophysics and Geochemistry, Developed under the Auspices of the UNESCO, Eolss Publishers, Oxford ,UK, p. 1-49.

Panza G.F., V. Kossobokov, A. Peresan, A. Nekrasova (2014). Why are the standard probabilistic methods of estimating seismic hazard and risks too often wrong? In: Earthquake Hazard, Risk, and Disasters. M. Wyss Ed., Chapter 12, 309-357. http://dx.doi.org/10.1016/B978-0-12-394848-9.00012-2.

Peresan A., A. Magrin, A. Nekrasova, V.G. Kossobokov, G.F. Panza (2013) Earthquake recurrence and seismic hazard assessment: a comparative analysis over the Italian territory. In: Proceedings of the ERES 2013 Conference. WIT Transactions on The Built Environment, Vol 132, pp 23-34. doi:10.2495/ERES130031, ISSN 1743-3509 (on-line)

Shedlock K.M., D. Giardini., G. Grünthal, P. Zhang (2000). The GSHAP Global Seismic Hazard Map. Seismol. Res. Letters 71 (6), 679-686.

Stucchi M, R. Camassi, A. Rovida, M. Locati, E. Ercolani, C. Meletti, P. Migliavacca, F. Bernardini, R. Azzaro (2007) DBMI04, il database delle osservazioni macrosismiche dei terremoti italiani utilizzate per la compilazione del catalogo parametrico CPTI04. Quad. Geof. 49: 38 (available at http://emidius.mi.ingv.it/DBMI04/).

Stucchi et al., (2012) The SHARE European Earthquake Catalogue (SHEEC) 1000–1899. Journal of Seismology, doi: 10.1007/s10950-012-9335-2.

Stein S., R. Geller, M. Liu (2012) Why earthquake hazard maps often fail and what to do about it, Tectonophysics, 562-563, 1–25.

Wyss M., A. Nekrasova, V. Kossobokov (2012) Errors in expected human losses due to incorrect seismic hazard estimates, Nat. Hazards, 62, 927–935.

Zuccolo E., F. Vaccari, A. Peresan, G.F. Panza (2011) Neo-deterministic and probabilistic seismic hazard assessments: a comparison over the Italian territory. Pure Appl. Geophys., 168:69–83.